# Hardware-accelerated Aggregation: Unification and Specialization

Alireza Shateri[T], Hongshi Tan[S], Michael Ng[T], Bingsheng He[S], Qizhen Zhang[T]

[T]University of Toronto, [S]National University of Singapore

## ABSTRACT

The high efficiency of domain-specific hardware has sparked substantial interest in adopting accelerators in data analytics systems. Among many choices, GPUs and FPGAs thrived as two popular solutions due to their prevalent deployments in cloud data centers. This paper investigates hardware acceleration solutions for aggregation, a critical data analytics operation. Specifically, we implement aggregation with a unified hardware acceleration framework, which trades efficiency for ease of programming and portability, and then further develop hardware-specific optimizations. We evaluate these solutions on three recent computing hardware platforms: a CPU, a GPU, and an FPGA, with metrics that cover both the performance and energy consumption of on-device and end-to-end processing.

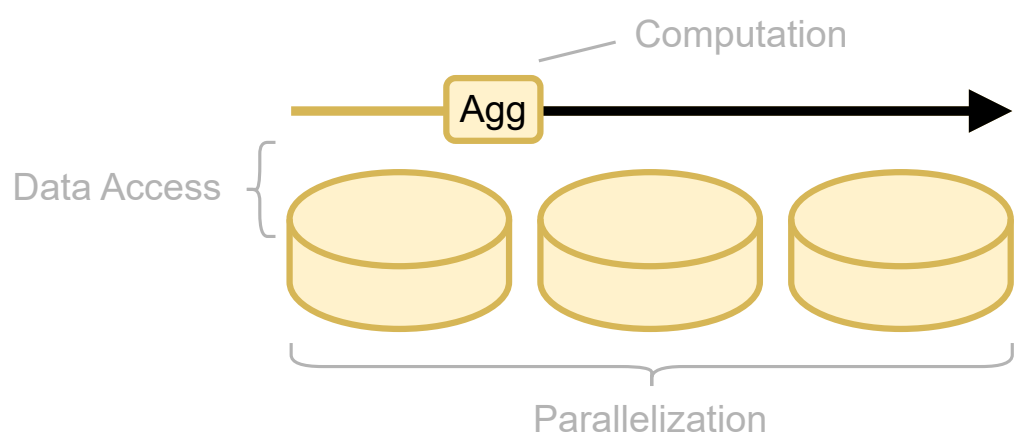

**Figure 1: Aggregation performance is determined by the efficiency of data access, parallelization, and function computation, stressing the underlying hardware in various aspects.**

## 1 INTRODUCTION

The demand for high compute efficiency in cloud applications continues to increase due to rapid data growth, but hardware scaling laws for general-purpose processors (i.e., CPUs) have slowed down over the years. Compared to CPUs, domain-specific accelerators, e.g., Graphics Processing Units (GPUs) and Field Programmable Gate Arrays (FPGAs), offer higher compute performance and energy efficiency. Accelerators have been widely deployed in cloud data centers to offload a variety of infrastructure and application functionalities [2–4, 9, 15, 27]. Generally, their high parallelism, e.g., tens of thousands of cores in recent GPUs, and memory speed, e.g., up to 1 TB/s bandwidth with High Bandwidth Memory (HBM), can drastically improve the performance of data analytics tasks. Prior work has shown promising results about utilizing FPGAs and GPUs for relational data analytics such as sorting [26, 32], projections [32], and joins [11, 12, 20, 31]

However, there is a lack of cross-hardware investigation into common relational data analytical operations. Different accelerators present different performance and energy characteristics. For instance, GPUs and FPGAs provide higher parallelism and on-device memory speed, but both accelerators clock less frequently than CPUs. A horizontal comparison between state-of-the-art hardware devices can provide meaningful insights.

In this paper, we present a study of the effect of hardware acceleration on data aggregation, an operation frequently involved in a variety of data analytics jobs, including MapReduce [7, 16], graph processing [22, 45], streaming and time series analytics [5, 13], and distributed machine learning [1, 17]. Despite many variants, the most fundamental task of data aggregation is to scan the data items in a given database and apply an aggregation function (e.g., `min` or `sum`), as shown in Figure 1. We focus on in-memory aggregation—otherwise the performance is dominated by storage I/O. Three factors determine the performance of in-memory aggregation: how fast data is accessed from memory, how efficiently the aggregation function is computed by the processing unit, and how well the operation is parallelized. To accelerate this operation with heterogeneous devices, we first use a unified hardware acceleration framework, called TornadoVM [10], which compiles a high-level implementation into binaries that can run on different target hardware via two levels of JIT compilation. This unification advocates ease of programming and portability but unfortunately sacrifices performance. We show the inefficiencies of the generated aggregation kernels by the framework and develop optimizations specialized for each hardware platform. Finally, we compare data aggregation performance between state-of-the-art CPU, GPU, and FPGA, to investigate the effect of hardware acceleration.

Our study involves metrics for both aggregation performance and its energy efficiency. In the analysis of the results, we consider not only on-device performance, i.e., data is preloaded into the device memory, but also end-to-end performance. The former represents an ideal scenario where the overhead of moving data between the host and the accelerator, which has been shown as the primary bottleneck of adopting hardware acceleration for databases [23, 43], is avoided. Including results for both setups paints a complete picture of the benefits and limitations of hardware acceleration.

This work makes the following contributions.

- Implementation of aggregation in a unified acceleration framework and hardware-specialized optimizations (§3).
- Evaluation of aggregation over realistic datasets on CPU (Intel Xeon) and popular data center accelerators: GPU (NVIDIA A100) and FPGA (AMD Alveo U55C) (§4).
- Cross-device comparison with metrics covering performance (§5) and energy consumption (§6).

## 2 BACKGROUND

### 2.1 Data Aggregation

Aggregation is a basic construct baked in many real-world data analytics systems, including cloud data warehouses [6, 34], Business Intelligence [24, 35], MapReduce [7, 16], graph processing [22, 45], streaming and time series analytics [5, 13], as well as distributed machine learning training [1, 17]. Optimizing its performance has practical meanings. Essentially, this operation scans the data items in a given database, applies an aggregation function (e.g., `min` or `sum`), and returns the summarized view of the data. As shown in Figure 1, aggregation stresses the underlying hardware in three aspects: how fast data items are accessed, how efficiently the aggregation function is computed, and how well the operation is parallelized. Prior work explored specific hardware, especially GPUs [18, 32], to improve aggregation performance, but there has been a lack of cross-hardware comparison and analysis, which can help to better navigate through the design space.

### 2.2 Hardware Acceleration

**Graphics Processing Units (GPUs).** GPUs are designed for high-throughput computation. They feature many simple cores that can execute thousands of threads in parallel. Figure 2 shows the high-level architecture of a modern GPU, which comprises multiple Streaming Multiprocessors (SMs). Each SM contains numerous Processing Elements (PEs) or cores that execute instructions in a Single Instruction, Multiple Data (SIMD) or Single Instruction, Multiple Threads (SIMT) fashion. The memory hierarchy in GPUs includes global memory, L2 cache, SM-local shared memory and L1 cache, and registers, each varying in size and access speed. With High Bandwidth Memory (HBM), even the slowest level, i.e., the global memory, in today's GPUs can provide terabits/s bandwidth.

**Field-Programmable Gate Arrays (FPGAs).** FPGAs have programmable logic blocks and gates (Figure 3). These blocks can be configured to implement functions tailored to specific applications. Blocks are interconnected via programmable wires, allowing users to further custom circuits. Modern FPGAs are equipped with sizable onboard memory (global memory). This memory is distributed across independent channels, each accessible by a memory engine (ME). Unlike fixed-architecture processors, FPGAs excel in parallel pipeline processing. Despite the promise, FPGAs are notoriously difficult to program. Designing an optimal hardware architecture requires deep expertise and careful resource management.

**Unified Framework.** TornadoVM [10] is a JVM plugin that allows Java applications to offload functions to heterogeneous hardware, such as multi-core CPUs, GPUs, and FPGAs, with a unified API. Specifically, its *task-schedule* abstraction specifies groups of Java methods as tasks that are automatically scheduled by the runtime on target hardware. To support heterogeneous hardware, TornadoVM adopts a two-tier Just-In-Time (JIT) compilation technique to specialize the code at compile time: Java bytecode from the application is first compiled to OpenCL [30] C code, which is further compiled to machine code by the OpenCL driver for the target hardware. Hence, TornadoVM serves as a unified framework that supports any OpenCL-compatible hardware. This framework has been adopted as a solution to transparently accelerate compute-intensive machine learning workloads in Flink with heterogeneous hardware [38]. Two annotations, `@Parallel` and `@Reduce`, are provided by TornadoVM for applications to parallelize loops and reduction operations.

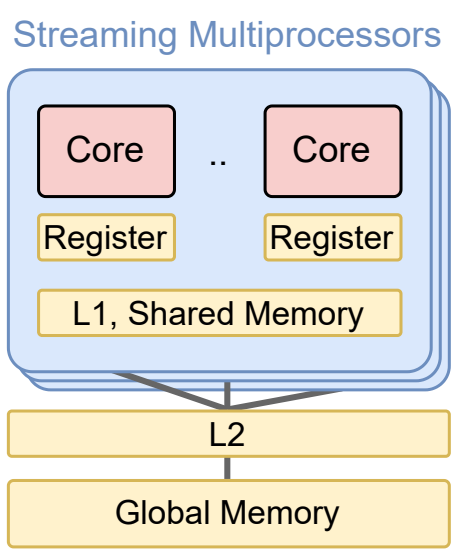

**Figure 2: GPU**

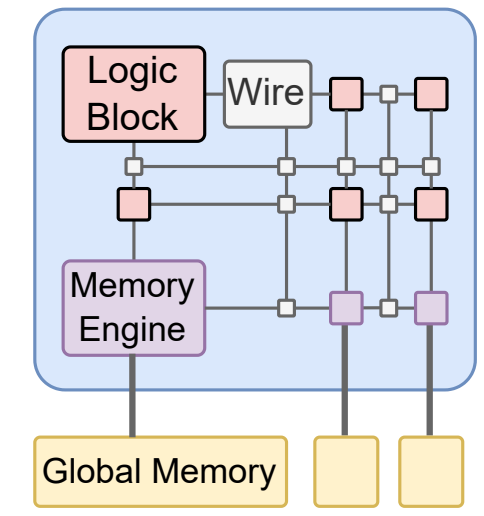

**Figure 3: FPGA (logics and wires are programmable)**

```java
class Aggregation {
    public void static kernel([Schema] input, @Reduce [
        Type] r) {
        for (@Parallel i = 0; i != input.size(); i++) {
            r.set(aggregate(r.get(), input.get(i)));
        }
    }
}
```

**Figure 4: Aggregation implemented with TornadoVM. This simple implementation is automatically compiled into hardware-local executables by the framework.**

## 3 ACCELERATION AND OPTIMIZATIONS

We first present aggregation implementation in the unified acceleration framework and then show how we improve its performance via hardware-specific optimizations.

### 3.1 Unified Acceleration

Figure 4 illustrates the aggregation implemented in Java with TornadoVM 's API. The kernel is a static Java method that iterates over input tuples to apply an aggregation function. We express data parallelism by annotating the loop with `@Parallel` and the output accumulator with `@Reduce`. The former executes each loop iteration in parallel, while the latter specifies that the result `r` is a reduction variable that must accumulate contributions from all iterations. Writing code in TornadoVM is similar to writing a regular Java program. With the annotations, the framework JIT-compiles the kernel to an intermediate OpenCL representation and further generates the executable for the target device.

**Inefficiencies.** While the unified approach provides ease of programming and portability, the automatically generated code does not exploit device-specific characteristics. The inefficiencies of the TornadoVM-generated aggregation per hardware are as follows.

- On CPUs, although multi-threading is exploited, by default TornadoVM configures thread count as the number of data items to aggregate. Spawning threads more than available cores incurs significant scheduling overhead and thread contention. In addition, within each thread, the reduction process linearly scans individual tuples. Advanced CPU features, e.g., SIMD instructions and cache locality, are not utilized.

- On FPGAs, the generic kernel sequentially reads input data from global memory and utilizes only one memory channel, while modern FPGAs often have multiple memory channels. Moreover, FPGA-specific spatial parallelism, e.g., unrolling loops to create pipelines, is not exploited and thus resources, including logic units and DSPs, are not fully utilized.
- On GPUs, TornadoVM generates two kernels for aggregation, one for computing the partial aggregate within each thread block and the other for summarizing the final aggregate. Global synchronization primitives, e.g., memory fence, are used to coordinate threads. Also, accesses to global memory are not coalesced and thus underutilize memory speed.

## 3.2 Specialization

**CPU Optimizations.** We first align the thread count with the number of available CPU cores. This eliminates the overhead of excessive threads. To utilize vectorized instructions, we adopt the `simd` directive in OpenMP to issue wide Intel AVX instructions in each thread to aggregate multiple tuples in one cycle.

**FPGA Optimizations.** Based on TornadoVM's unified acceleration framework, the high-level aggregation kernel is compiled into a hardware data processing element (PE) that is instantiated directly on the FPGA fabric. To fully utilize the memory bandwidth from the FPGA's multiple memory channels, we parallelize aggregation across multiple PEs, each concurrently processing a partition of the input data. As shown in Figure 5, given an FPGA equipped with $N$ memory channels, the input tuples are evenly partitioned across $N$ channels. Each channel is assigned a dedicated read engine, which fetches tuples from global memory and streams them into the processing pipeline for data aggregation.

To enhance performance and better exploit FPGA-specific architectural features, we further incorporate two key building components into the generated design. First, we optimize the scalar memory loads generated from TornadoVM with 512-bit vector reads to saturate the external memory bandwidth and ensure that each PE receives data at full throughput. These accesses are independently mapped to separate memory controllers, enabling spatial parallelism across banks without arbitration or cache contention. Second, we incorporate the aggregation computation through a statically scheduled reduction tree fully realized in registers. Each input stream is processed by a pipelined adder network with an initiation interval of one cycle, allowing the production of one aggregated result per cycle. In contrast to the default TornadoVM-generated design, which relies on atomics and memory fences to resolve conflicts, our optimization guarantees deterministic and conflict-free updates by construction. This eliminates synchronization overhead and allows the pipeline to operate at full utilization.

**GPU Optimizations.** We fuse the two GPU kernels into a single one and exploit *cooperative groups* (CGs) [29] to improve synchronization and memory efficiency. Specifically, thread count is aligned with the number of SMs on the device and accesses to the input tuples from the same thread block are coalesced with *grid-stride loops* [28]. At runtime, threads perform block-local reductions by accumulating over tuples entirely in registers. Finally, partial aggregates of blocks are merged with a global synchronization.

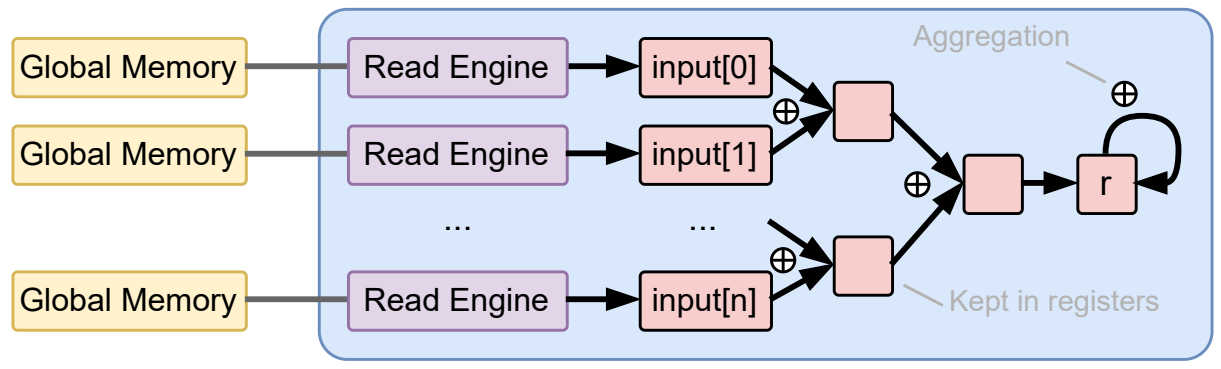

**Figure 5: Optimized aggregation on FPGAs.**

# 4 EXPERIMENTAL SETUP

**Workloads and Metrics.** The dataset where we run aggregation is the `lineitem` table (`sum` over the `suppkey` attribute) from the TPC-H benchmark with varying scale factors. We evaluate two scenarios: data resides in the device memory (where there are no data transfers between the host and the device), and data resides in the host memory. In addition to aggregation performance, we also consider energy consumption (in Joules) to provide more perspectives.

**Hardware.** Our experiments are conducted on recent production hardware. We run CPU experiments on a server equipped with an Intel Xeon Silver 4314 CPU with 16 physical cores (32 hyperthreads), each clocking at 2.4 GHz, and 128 GB of DDR4 memory. For FPGA experiments, we use an AMD/Xilinx Alveo U250 FPGA, which offers 1.7 million logic look-up tables and 12 K DSP slices and 64 GB of four-channel DDR4 memory, offering up to 77 GB/s of sequential access bandwidth. Finally, the GPU experiments are carried out on an NVIDIA A100 GPU, which consists of 6912 CUDA cores and 432 Tensor cores and 80 GB of high-bandwidth HBM2 memory.

# 5 PERFORMANCE

We first present the aggregation speed achieved by the unified framework on each of the three hardware devices and then show the effectiveness of our hardware-specific optimizations. In addition to on-device performance, we also report end-to-end results that account for data transfer times between the host and the accelerator.

## 5.1 Unified Acceleration

The first-hand aggregation performance of TornadoVM on TPC-H scale factor 200 is shown in Figure 6. The CPU takes 0.5 seconds to aggregate 9.6 billion tuples—19 billion tuples/s. The FPGA is significantly slower with 30 million tuples/s. The GPU achieves the highest performance at 660 billion tuples/s, 35× and 19000× faster than the CPU and the FPGA, respectively. This experiment shows that although the unified framework can achieve highe performance the GPU, it generally underutilizes hardware capability for aggregation due to the inefficiencies we identified (§3).

## 5.2 Optimizations

Our hardware-specific optimizations can effectively improve aggregation performance. As Figure 6 shows, our first CPU optimization, which aligns thread count with available cores, brings 1.7× speedup.

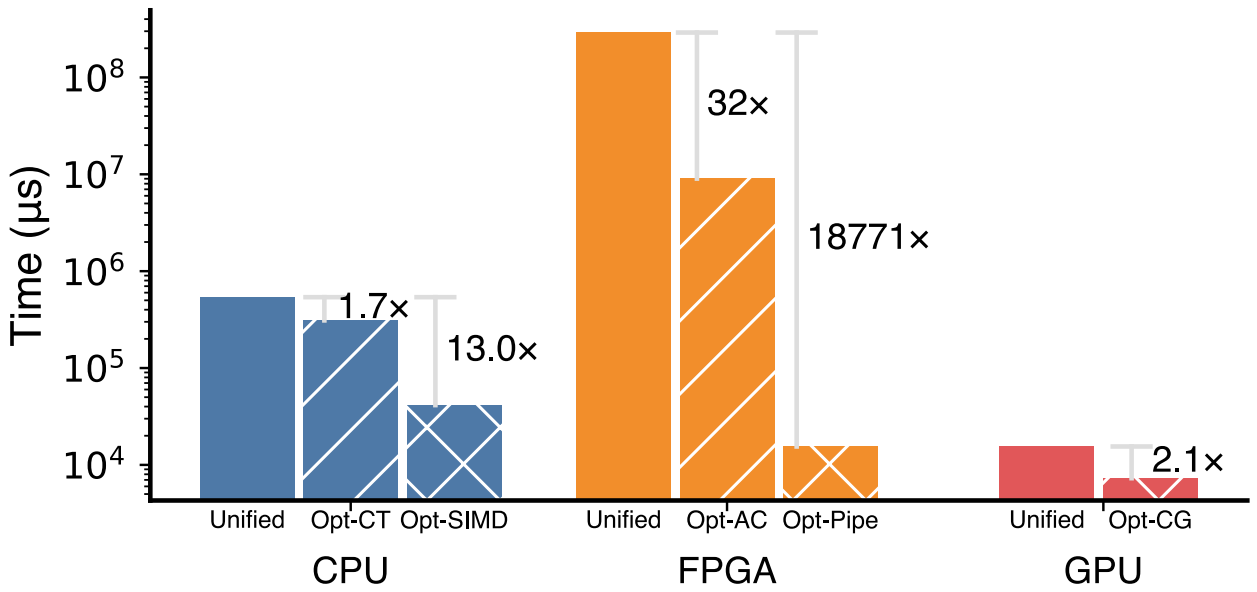

**Figure 6: Unified and specialized aggregation performance on different hardware. CT, SIMD, AC, Pipe, CG refer to our optimizations for CPU (configured threads, SIMD), FPGA (all channels, optimized pipeline), and GPU (cooperative groups).**

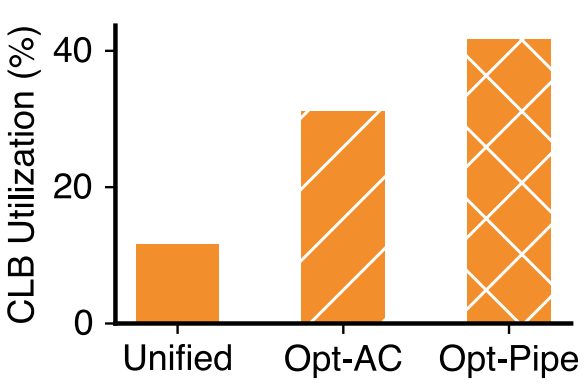

**Figure 7: CLB utilization**

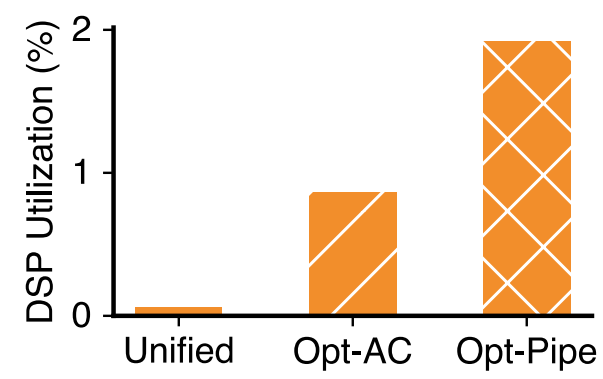

**Figure 8: DSP utilization**

With SIMD instructions, CPU aggregation is 13× faster than the unified code. The improvement is most pronounced for FPGA execution: utilizing all memory channels improves speed by 32×, and FPGA performance is fully unlocked with our optimized pipeline, which brings *five orders of magnitude* improvement. The FPGA is now 2.7× faster than the CPU. Figures 7 and 8 explain why our FPGA implementation is more efficient: FPGA resources, i.e., Configurable Logic Block (CLB) and Digital Signal Processing (DSP) units, are much better utilized. For GPU execution, although TornadoVM can already utilize this accelerator to achieve higher performance than the CPU, our optimization with cooperative groups can further elevate GPU aggregation speed by 2.1×.

Figure 9 shows scalability of optimized aggregation with the data size, from scale factor 50 to 200. The trend is clear: the GPU is the fastest hardware in all scenarios, followed by the FPGA. They achieve 5.7× (1.3 trillion tuples/s) and 2.7× (620 billion tuples/s) higher throughput than the CPU (232 billion tuples/s), respectively.

These results show that (1) our hardware-specific optimizations effectively eliminate the inefficiencies in the unified framework, and (2) hardware accelerators, when fully exploited, can provide significantly faster aggregation than CPUs.

## 5.3 End-to-end Comparison

We finally show the end-to-end perspective of aggregation performance, which accounts for the times of moving data from the host memory to the accelerator's device memory. Figure 10 reports the results with hardware-specific optimizations on different scale factors. Data transfers between the host and the accelerator via PCIe (v4.0) are the bottleneck: they contribute to 97.9% of the GPU execution time and 96.8% of the FPGA execution time. Hence, the throughputs of the GPU (61 billion tuples/s) and the FPGA (45 billion tuples/s) are significantly lower than that of the CPU (232 billion tuples/s). This result shows that even with advanced hardware today, to benefit data analytics with hardware acceleration the host-device data transfer bottleneck must be addressed.

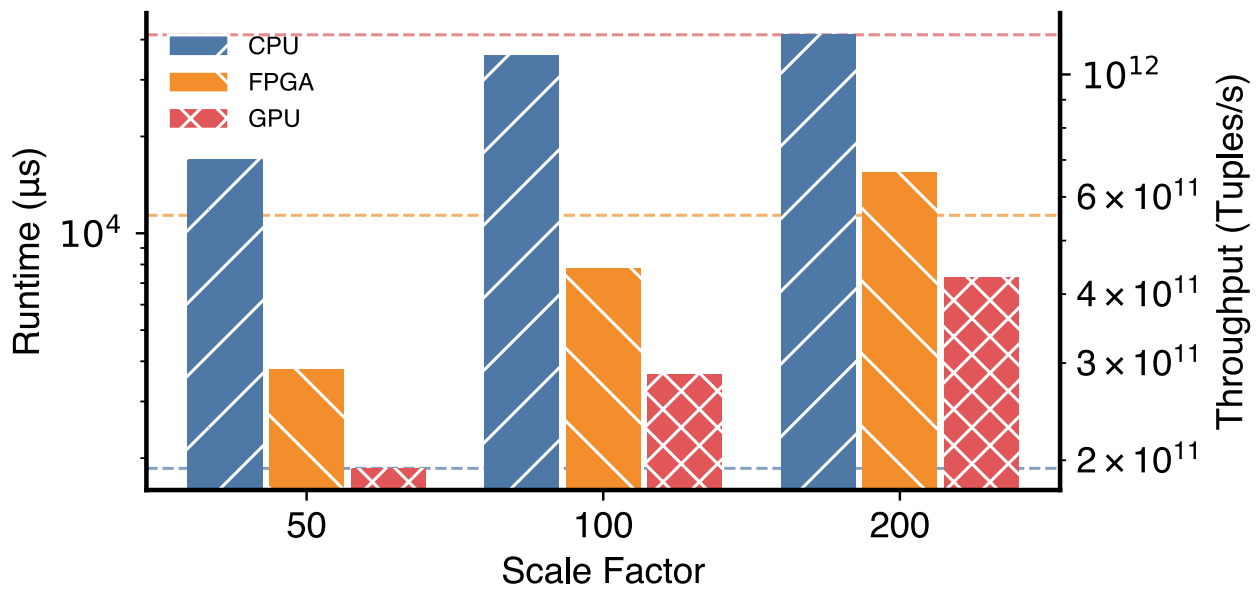

**Figure 9: Optimized aggregation performance with varying data sizes. Dashed lines represent the throughput of GPU, FPGA, and CPU (top to bottom).**

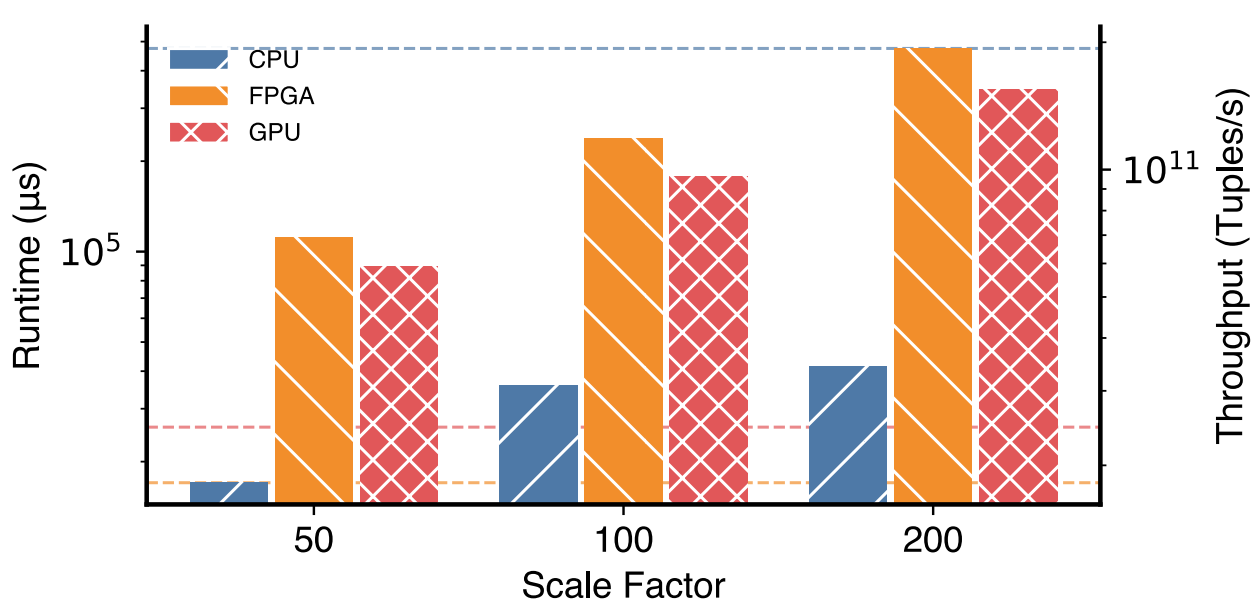

**Figure 10: End-to-end performance. Dashed lines represent the throughput of CPU, GPU, and FPGA (top to bottom).**

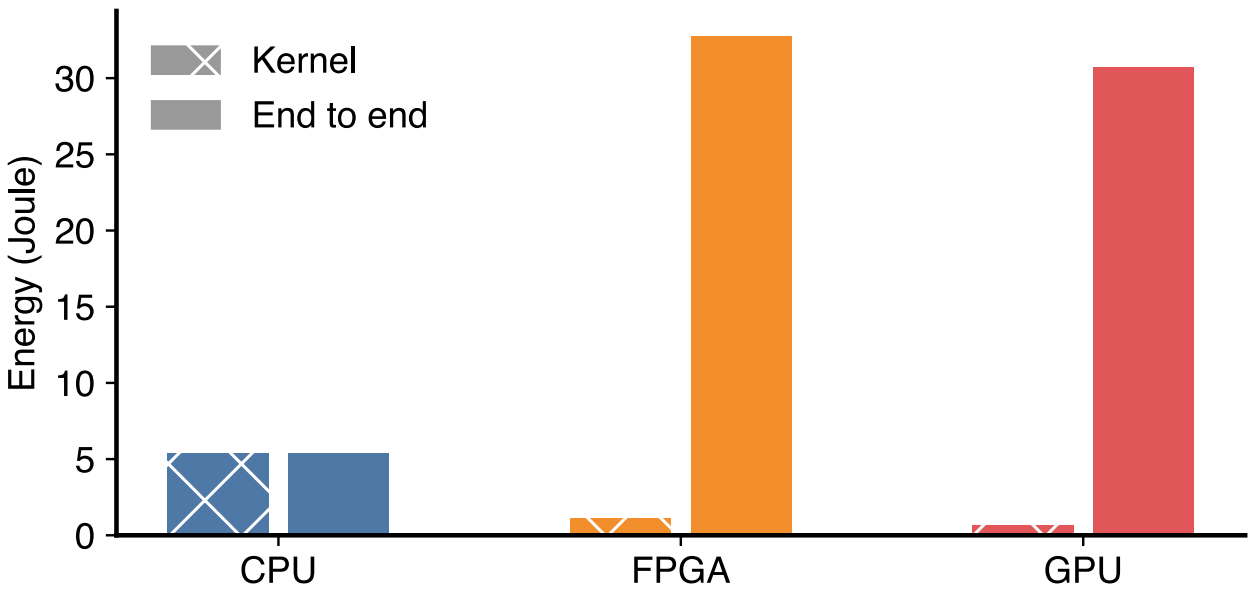

**Figure 11: Energy consumption on different hardware.**

## 6 ENERGY CONSUMPTION

**Kernel Power Consumption.** When performing on-device aggregation, hardware accelerators are not only faster but also more energy-efficient. Figure 11 shows that the CPU consumes 5.5 Joule (J) to aggregate 9.6 billion tuples (scale factor 200). In comparison, the FPGA consumes only 20% (1.1 J) and the GPU consumes only 10% (0.6 J) of the CPU power usage to aggregate the same tuples.

**End-to-end Power Consumption.** Similar findings in performance comparison also apply to energy consumption. When data transfers are included, the hardware accelerators consume significantly more power than the CPU. Specifically, as the CPU aggregates the data *in situ*, its power consumption remains 5.5 J. The GPU is slightly more efficient than the FPGA, but both accelerators consume more than 30 J to complete the end-to-end processing. This result confirms that data transfers are the primary road blocker for developing efficient hardware solutions for aggregation.

# 7 RELATED WORK

**Aggregation on GPUs and FPGAs.** The primary focus of recent work about accelerating aggregation with GPUs and FPGAs has been on group-by aggregations based on sorting or hashing. Specifically, on GPUs, Karnagel et al. [14] implemented a GPU architecture for hash-based group-by and tuned configuration parameters to improve performance. Luan et al. [18, 19] studied CPU-GPU co-processing via inter-operator parallelization to optimize hash-based group-by and aggregation together. More recently, Wu et al. [36] investigated the inefficiencies of GPU execution for join-then-aggregate queries and proposed optimizations.

On FPGAs, Zhang et al. [44] implemented various hash functions on FPGAs and use them to execute hash-based group-by. Moghaddamfar et al. [25] designed group-by aggregation with FPGA on-chip cache and pipeline for enterprise DBMS. Xue et al. [40] developed a fully streaming FPGA pipeline for sorting-based group-by aggregation. Eryilmaz et al. [8] proposed a CPU-FPGA architecture for aggregation and a model that predicts the processing time.

In comparison, we investigate the impact of unified and specialized acceleration solutions on the foundational aggregation primitive and perform cross-hardware comparison. We also analyze the energy efficiency of different solutions.

**Unified hardware acceleration.** Xekalaki et al. [39] adopted TornadoVM in Flink to accelerate various ML tasks with FPGAs and GPUs. We focus on aggregation and develop hardware-specific optimizations on top of the unified framework.

# 8 CONCLUSION AND FUTURE DIRECTIONS

We investigated the benefits of popular data center hardware accelerators, i.e., FPGAs and GPUs, on data aggregation, compared to CPUs with performance and energy metrics. Our key contributions include implementing aggregation with a unified acceleration framework (TornadoVM), developing hardware-specific optimizations, and evaluating these solutions on recent hardware. These results are useful for developing efficient aggregation solutions.

Our findings encourage the following directions for future explorations.

**Transparent Specialization.** Our results reveal the trade-off between ease of programming and efficiency in the current unified hardware acceleration framework. The fully hardware-specialized optimizations are manually implemented and separate from TornadoVM. It has a strong motivation to reflect the specialized optimizations in a unified framework to achieve both properties. However, doing so is challenging without heavy system modifications. A future direction is to incorporate hardware-specific characteristics within the TornadoVM's backend, thereby moving specialization from applications to the framework.

**Extended Specialization.** The current specialized code targets aggregation, e.g., SIMD reduction on the CPU and lockless reduction tree on the FPGA. In the future, we plan to expand these optimizations for more relational data analytical operators. Although some techniques, e.g., configured threads and fully-utilized memory channels, are applicable to all operators, other techniques, e.g., vectorized reduction, need adaptation for operators other than aggregation. We will also consider opportunities for fusing different operators [37] on different hardware platforms, which is crucial for more complex queries.

**End-to-end Optimizations.** Despite orders of magnitude higher performance and energy efficiency achieved in the kernel execution for the FPGA and the GPU, these accelerators are currently outperformed by the CPU due to expensive data transfers between the host and the device. We expect the cost to increase for other relational operators, e.g., joins, where more output data is moved from the accelerator to the host. This road blocker must be addressed to benefit data analytics from hardware acceleration. An important future direction is to optimize data transfers between the host and the device. Existing work [21, 33] has exploited faster host-device interconnects and compression to speed up data transfers for certain operations, which are promising starting points.

**At-scale Unification.** TornadoVM has limited support for multi-device execution. As hardware accelerators are increasingly deployed in data centers, multi-accelerator data analytics are likely to become a commonplace in the future [41, 42]. We plan to investigate cross-device optimizations for large-scale data processing.